\documentstyle[preprint,aps,amssymb]{revtex}
\begin{document}
\draft
\title{Crystallization in a model glass:\\
influence of the boundary conditions}

\author{Philippe Jund and R\'emi Jullien}

\address{Laboratoire des Verres, UMR 5587 CNRS,
      Universit\'e Montpellier II, Place Eug\`ene Bataillon,
                     34095 Montpellier Cedex 5, France}

%\date{\today}
\maketitle

\begin{abstract}

Using molecular dynamics calculations and the Vorono\"\i\ tessellation,
we study the evolution of the local structure of a soft-sphere glass 
versus temperature starting from the liquid phase at different 
quenching rates. This study is done for different sizes and for two
different boundary conditions namely the usual cubic periodic boundary 
conditions and the isotropic hyperspherical boundary conditions for 
which the particles evolve on the surface of a hypersphere in four dimensions.\\
Our results show that for small system sizes, crystallization can indeed be
induced by the cubic boundary conditions. On the other hand we show that 
finite size effects are more pronounced on the hypersphere and that 
crystallization is artificially inhibited even for large system sizes. 

\end{abstract}

\eject

When cooling a liquid down to low temperatures, what is the nature of
the resulting solid phase ? The answer to this question 
in experimental studies or in
numerical studies depends on how fast the liquid has been cooled down i.e.
on the quenching rate. If the liquid is cooled faster than a critical rate, 
the resulting phase will be a glass otherwise the system will evolve towards
the more stable crystal phase. The knowledge of this critical rate is 
very important especially in numerical studies of model glasses 
since crystallization effects should not bias the results at low temperature. 
Similarly to what had been done in a Lennard-Jones system \cite{Nose} or in 
liquid sodium \cite{Wata} we used the Vorono\"\i\ cell statistics to determine
this critical rate in a model soft-sphere glass \cite{prl97}. Nevertheless
in numerical studies due to computer time and memory limitations, one has 
to mimic macroscopic samples with the use of the so-called 
``boundary conditions''. This is usually done by replicating a cubic box
containing a few thousands of particles, for which the calculations
will be explicitly done, in the 3 directions of space \cite{tildes}. 
These boundary conditions are generally called periodic boundary 
conditions (PBC) and let us call this periodic space $C_3$. Obviously 
in this geometry the finite systems stay homogeneous but can bear a slight 
anisotropy. Therefore it is justified to address the question of the
influence of this anisotropy on the crystallization of our model glass for
which the ground state is a face-centered-cubic arrangement. In
other words, do the cubic PBC favor crystallization ?\\
In this letter we investigate the influence of the boundary conditions and
the system size on the evolution of the structure of a soft-sphere
system during the quench with the use of classical 
molecular dynamics (MD) simulations. To distinguish an amorphous 
structure from a crystal or a partially crystallized system, we use the 
fraction of pentagonal faces of the Vorono\"\i\ cell attached to each particle, which is an excellent tool to obtain this distinction \cite{prl97}. As an 
alternative to the cubic PBC we use the ``hypersphere boundary conditions''
. This means that the MD simulations were performed by confining the
system of soft-spheres to the surface $S_3$ of a hypersphere in four dimensions.
This technique was used earlier for systems of charged particles \cite{caill},
or fluids of hard-spheres \cite{tobo} and Lennard-Jones particles \cite{shrei}.
The space $S_3$ (as well as $C_3$) is homogeneous, finite and non-Euclidean; but in addition, $S_3$ is isotropic therefore there is no preferential direction
which could influence crystallization. 
Moreover for the calculation of long-range forces like Coulomb forces 
an advantage of using $S_3$ is a 
reduction of computer time by a factor 3-4 compared to the 
standard Ewald method \cite{prlcaill}. Of course asymptotically for very
large system sizes the results in $C_3$ and $S_3$ should converge.\\
Our results show that for small system sizes crystal nucleation occurs
in $C_3$ while the system remains amorphous in $S_3$. This is coherent
with previous results obtained for hard-sphere collections \cite{tobo} and
shows that the glass stability is, as desired, higher on the hypersphere.
Nevertheless for larger systems we show that, compared to the asymptotic
behavior which is as expected identical in $C_3$ and $S_3$ , the use
of hyperspherical boundary conditions can artificially discourage crystal
nucleation and therefore the value of the critical quenching rate can
be underestimated. These results permit to shed new light on the influence
of the widely used PBC on crystallization effects in model glasses and
show the consequences of using isotropic boundary conditions in a curved
three dimensional space such as $S_3$.\\

In our microcanonical MD simulations the particles interact via the repulsive
potential proposed by Laird and Schober \cite{scho}:

\begin{eqnarray}
U(r) & = & \epsilon({\sigma\over r})^6 + Ar^4 + B \hbox{\ \ for\ \ } r  <  
3\sigma \label{line1}\\
     & = &  0  \hbox{\ \ for\ \ } r \ge 3\sigma \nonumber
\end{eqnarray}
The constants $A$ and $B$ are chosen so that both the potential and the
force are zero at the cut-off distance $3\sigma$. To avoid reduced units, 
we use the values of
Lennard-Jones argon for the mass $m$ of the particles, 
as well as for $\epsilon$ and $\sigma$ \cite{gazzi}. To integrate the 
equations of motion we
use the Velocity Verlet algorithm \cite{tildes} with a timestep 
$\Delta t =2.5$ fs. To investigate the influence of the system size we use 
samples containing $N=512$, $1000$ and $5000$ particles and to get 
better statistics we use either 10 independent configurations for 
$N=512$ and $1000$ or 2 configurations for $N=5000$. For each system we
start from a well equilibrated liquid sample at a temperature around
50K (well above the melting temperature $T_m \approx 23$K \cite{hoover}) which is then cooled down to zero 
temperature at two different quenching rates: $4.0\times10^{11}$K/s and
$10^{11}$K/s. The fastest quenching rate is close to 
the critical rate for which no crystallization should be observed after
the quench while the second rate is smaller then this critical rate and 
therefore the system should be in a crystallized or partially crystallized
state at the end of the cooling process \cite{prl97}. To study the 
evolution of the structure as a function of temperature we save all along
the quenching procedure configurations (positions and velocities) on which
the Vorono\"\i\ tessellation scheme is applied (it is worth noticing that 
no relaxation period has been considered). Once the Vorono\"\i\
cell attached to each particle has been determined, we can calculate 
the fraction $f_5$ of pentagonal faces in the system. This is a good
indicator of the degree of randomness (or on the other hand the degree of
crystallization) since a large value of $f_5$ (typically around 0.45) is
a sign of strong icosahedral local order characteristic of a glass phase, 
while a small value of $f_5$ ($< 0.2$) indicates crystal nucleation. Therefore
we know for each sample if crystallization took place during the quench and
we can determine at which temperature this crystallization started. \\
While MD simulations in $C_3$ are common and well established, we think
it useful to give more details on the simulations in $S_3$. First of
all concerning the density $\rho$, it is fixed and we use the 
value 
$\rho = N\sigma^3/V_{space} = 1$.
In $C_3$ we use standard PBC with a rigid cubic box of 
edge length $L$ ($V_{space}=L^3$)
while in $S_3$ $V_{space} = 2\pi^2R^3$ where $R$ is the radius of the 
hypersphere. Fundamentally the volume fractions of the corresponding 
hard-sphere systems are not strictly equal in both spaces because 
one should take into
account the curvature of the hard-spheres in the fourth dimension. Nevertheless
a straightforward calculation shows that the relative error in $S_3$ 
is smaller than $5\times10^{-3}$ \% for the smallest system considered 
here ($N=512$). Second let us detail a MD step 
in $S_3$. For convenience we chose to use the quaternion
formalism \cite{quater} in order to describe the equations of motion in 
the four dimensional space. 
The unit quaternion $Q^i=(q_0^i,q_1^i,q_2^i,q_3^i)$ attached to particle $i$ 
is defined by the set of $q_k^i$s given by
$q_k^i = x_k^i/R$,
where the $x_k^i$s are the coordinates of particle $i$ in four dimensions, 
verifying $\sum_k{(x^i_k)^2 = R^2}$. 
We consider the potential
$U(r)$ as being a function of the interparticle cartesian distance $r$ 
in the four dimensional
space. Thus the force $F^i$ on particle $i$ is a four dimensional vector, 
defined by the components $f_k^i$ given by: 
\begin{equation}
f_k^i = \sum_j{\frac{(q^j_k - q_k^i)}{|Q^{ij}|}}
\left( \frac{\partial U}{\partial r} \right)_{\rm r=r_{ij}}
\end{equation}
where $|Q^{ij}|$ is related to the distance $r_{ij}$ between particles $i$ and $j$ by $|Q^{ij}| = r_{ij}/R$. Next the force is
projected on the three dimensional  tangent space of particle $i$ by 
calculating the quaternion product \cite{quater}
\begin{equation}
\Phi^i = \tilde Q^i F^i
\end{equation}
where $ \tilde Q^i = (q_0^i,-q_1^i,-q_2^i,-q_3^i)$ is the 
``conjugate'' of $Q^i$. 
The acceleration of particle $i$ in the three dimensional tangent 
space $\gamma^i_k$ is obtained using the Newton's equations, 
\begin{equation}
\gamma^i_k = \Phi^i_k/m \hbox{\ \ ,\ \ } k=1,2,3
\end{equation}
Once the acceleration has been calculated, the velocity is recovered using
the standard Velocity Verlet scheme \cite{tildes} and then the particle $i$
can be moved to its new positions ${Q^i}^\prime$ using
\begin{equation}
{Q^i}^\prime = \frac{\delta Q^i Q^i}{|\delta Q^i|}
\end{equation}
with $ \delta Q^i = (1,\delta q_1^i,\delta q_2^i,\delta q_3^i)$ and
$\delta q_k^i = (\Delta tv_k^i + (\Delta t^2/2)\gamma_k^i)/R$. 
The determination
of the new positions completes the iteration and the process can be repeated
as desired. In the scheme described above the dynamics is  in fact 
performed in the four dimensional space with the constrain that the points
should stay on the hypersphere. This implies that the first component of the
force $f_0^i$, orthogonal to the hypersphere, is compensated by an adequate 
reaction force. We could have considered an alternative procedure consisting 
in staying on the hypersphere and considering forces between particles 
separated by the ``geodesic'' distance\cite{caill}. But, since only 
distances smaller than $3\sigma$ are considered 
here (for larger distances the potential is zero), 
the relative difference (due to the curvature of the hypersphere)
between the cartesian distance and the geodesic distance 
remains smaller than a fraction of a percent for the smallest system 
considered here (512 particles). Therefore the quantitative
differences between the two approaches stay very small and in fact
vanish in the infinitely large system limit.\\
Finally concerning
the Vorono\"\i\ tessellation in $C_3$, the scheme has been detailed elsewhere
\cite{molsim}. In $S_3$ the tessellation is done in the four dimensional space
using the geodesic distance between particles \cite{jsm}.
For all the different samples and quenching rates, all the calculations have
been made in parallel in $S_3$ and in $C_3$ in order to compare the effect
of the boundary conditions on the final structure of the samples after
the quench.\\

First we consider the fastest quenching rate ($4.0\times10^{11}$K/s).
In Fig. 1 we report on the variation in $C_3$ and $S_3$ of $f_5$ versus the 
temperature for the three sizes considered in our study ($N=512, 1000, 5000$).
It is worth recalling that the results have been averaged over several samples
for the different values of $N$. The first observation concerns the asymptotic
behavior obtained for the largest systems in both spaces: as expected this behavior is similar. Indeed if $L$ or $R$ increases, the
description becomes closer to the one of the infinite system.             
As a general result, the finite size effects are more important 
for temperatures below the melting temperature. 
Moreover, if one excludes the case $N=512$ in $C_3$, it appears that these 
finite size effects are more important in $S_3$ than in $C_3$. 
This is clearly visible when comparing the curves for
 $N=5000$ and $N=1000$ which are almost superimposed in $C_3$ and 
not in $S_3$.
These finite size corrections act in opposite directions 
when comparing $C_3$ and $S_3$: in $C_3$ the small system sizes favor a small
value of $f_5$ while in $S_3$ high values are favored. Indeed, 
the smaller the hypersphere, the higher $f_5$ i.e. the more the system 
is icosahedral. This can be explained by the fact that for
$N=120$ the ground state in $S_3$ is the ``Polytope \{3,3,5\}'' \cite{poly}
made of dodecahedral Vorono\"\i\ cells only, which means that for
this ideal structure $f_5=1$. Therefore the more $N$ decreases and comes
closer to $120$ the more the structure will resemble the Polytope \{3,3,5\}
and the more $f_5$ will increase and tend towards 1.
The fact that even for large systems $f_5$ remains greater than 0.43 at T=1K, 
indicates that no crystallization occurred during the quench and shows that
the quenching rate is above the critical rate separating glass and crystal
forming rates. Another point concerns the glass transition which occurs
around $T_g \approx 10.5$K \cite{scho} and which is reflected in a saturation
of $f_5$ below $T_g$. This saturation is not obvious for $N=512$ but is
already visible for $N=1000$ and is much more pronounced for $N=5000$. 
Our results do not show a significant shift of $T_g$ as a function of 
system size.\\
The huge difference between $C_3$ and $S_3$ is observed
for $N=512$. Indeed in $C_3$, for $N=512$, Fig. 1 shows clearly a 
drop of $f_5$ below  $T_g$ with a saturation towards $0.36$ at 
low temperature. Since these results have been averaged over ten samples,
 this indicates that some of the samples have crystallized during 
the quench. Snapshots of the structures at T=1K, confirm that 
three out of ten samples are crystals, one of them having a 
fraction of pentagonal faces smaller than
$10^{-2}$ indicating an almost perfect crystalline character.
These results are coherent with what has been observed in hard-sphere
collections \cite{tobo} and indicate that for small systems 
(a few hundred particles) the cubic PBC can indeed induce crystallization
while the same systems remain ``perfectly'' amorphous with hyperspherical
boundary conditions. Nevertheless this ``encouragement'' to crystal
nucleation in $C_3$ already disappears for systems containing 
$1000$ particles which is the size we have used in earlier 
studies \cite{prl97}.\\
In a second step we consider a quenching rate of $10^{11}$K/s. This rate
is smaller than the critical rate and therefore the structure after the 
quench should be crystalline. In Fig. 2 we show the variation 
of $f_5$ versus the temperature in $S_3$ and $C_3$ for $N=1000$ 
and $N=5000$. As expected the asymptotic behavior ($N=5000$) is the same 
in both spaces: a drop of $f_5$ is observed below $T_g$ indicating 
crystallization. In $C_3$ this effect is even more spectacular for the 
systems containing a thousand particles. In fact almost all of the ten 
samples considered in this study show a value of $f_5$ smaller than 
$0.2$ at T=1K indicating that these samples have undergone crystallization.
A look at snapshots of these crystalline structures does not show an 
obvious correlation between the axes of the periodic cubic box and the
crystallographic directions. The fact that the samples containing 
$N=1000$ particles exhibit an almost perfect crystalline character may 
be due to the difficulty to build extended defects in such small 
crystalline systems. This is not the case for the 
$5000$ particles samples which may explain why these samples contain more
defects and therefore show a higher value
of $f_5$ at small temperature ($f_5 \approx 0.35$). Nevertheless crystal
nucleation occurred in $S_3$ and $C_3$ for $N=5000$ and in $C_3$ for
$N=1000$. On the contrary
the systems containing a thousand particles in $S_3$ do not show any sign
of crystal nucleation. This indicates that for small system sizes the
hyperspherical boundary conditions discourage {\em artificially} crystal
nucleation. If the aim is to avoid crystallization then one should indeed 
work in $S_3$ with a relatively small system. On the other hand if one 
wants to determine the critical rate separating crystal forming rates 
from glass forming rates then the use
of hyperspherical boundary conditions can lead to an underestimation 
of this rate. Indeed we quenched liquid samples containing $1000$ particles
at a rate of $4 \times 10^{10}$K/s and still we did not detect any sign of
crystallization in $S_3$ even though the $5000$-particles system crystallized.
This is another example of the strong size effects in $S_3$: one needs to use
a large system to get the correct results. On the contrary in $C_3$ the
correct behavior is already obtained for systems containing $1000$ particles.\\

In conclusion, we have used molecular dynamics simulations and the 
Vorono\"\i\ tessellation to study the influence of the boundary
conditions on the structure of the solid phase obtained after
the quench of a soft-sphere system. This has been done for
several system sizes ($512, 1000$ and $5000$ particles) 
and for two different quenching rates close
to the critical rate separating crystal from glass forming rates.
To simulate a macroscopic sample we have used the usual cubic
periodic boundary conditions (PBC) but also isotropic hyperspherical boundary
conditions for which the particles are confined to the surface of 
a hypersphere in four dimensions. For the glass forming rate we 
show that for the smallest system size, the PBC can indeed 
induce crystal nucleation while the larger samples remain amorphous. 
This shows that caution should be used with the usual PBC when 
samples containing only a few hundred particles are considered. On the 
contrary for the crystal forming rate the
results show that the hyperspherical boundary conditions can discourage
{\em artificially} crystal nucleation even for samples containing
thousand particles. On the one hand this is an advantage because 
no crystallization effects bias the results supposed to be obtained in 
a glass. On the other hand for a given system it leads to an underestimation
of the critical quenching rate. 
Of course the asymptotic results obtained with $5000$ particles are
as expected similar using both boundary conditions but since it appears
that the finite size effects are smaller when the usual PBC are used and
even though the calculations on the hypersphere are slightly faster, our
results indicate that in any case the correct behavior is obtained with
the usual PBC for the 1000-particles samples. Of course this critical size
depends on the interaction potential and should be re-determined for each 
different system.

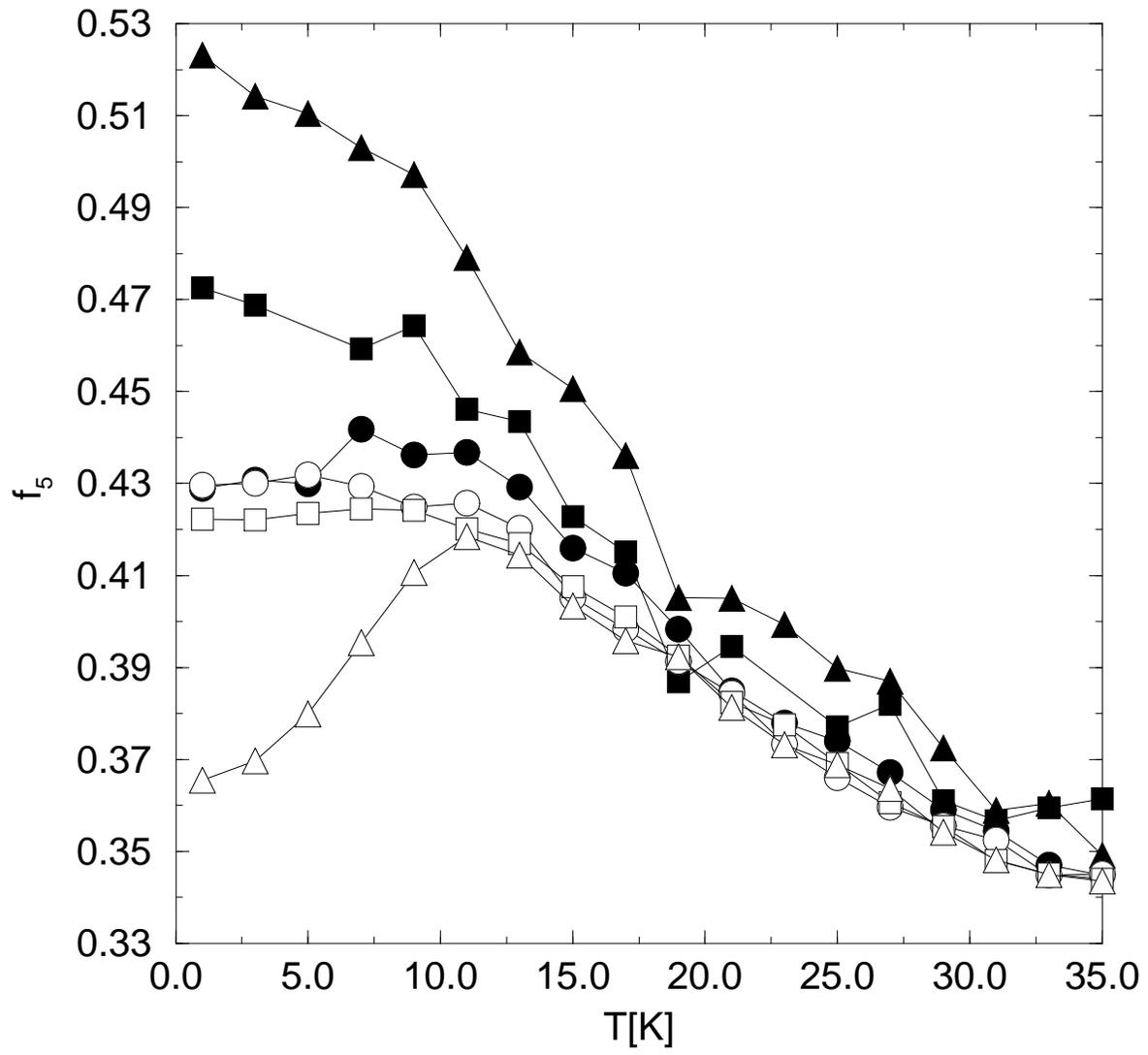
\begin{figure}
\caption{Fraction of pentagonal faces, $f_5$, versus temperature in 
$C_3$ (open symbols) and $S_3$ (black symbols) for $N=512$ (triangles),  $N=1000$ (squares) and $N=5000$ (circles).\\
The quenching rate is equal to $4\times10^{11}$K/s.     
}
\label{Fig1}
\end{figure}

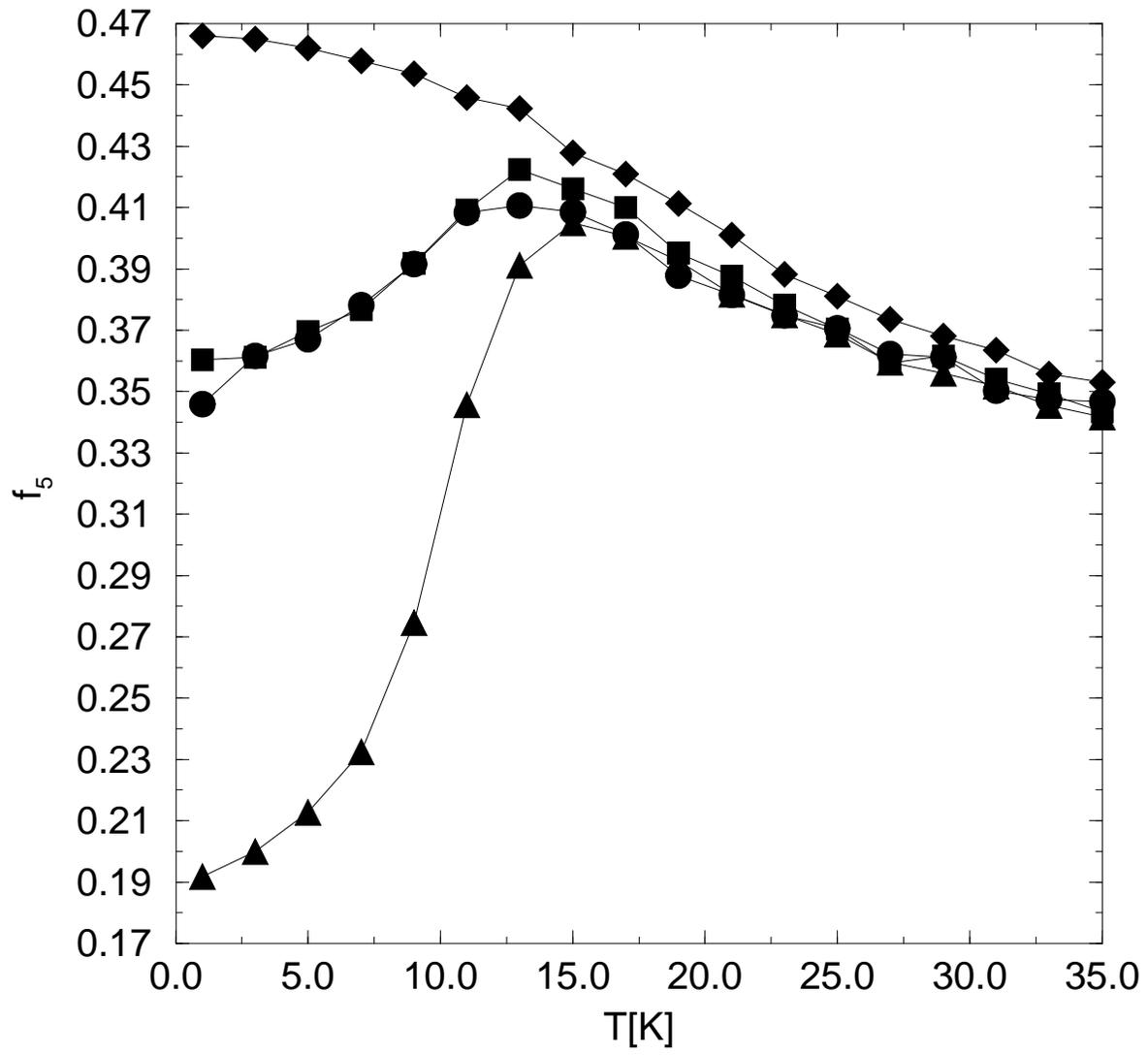
\begin{figure}
\caption{Fraction of pentagonal faces, $f_5$, versus temperature in 
$C_3$: $N=1000$ ({\Large $\blacktriangle$}) \\ and $N=5000$ ({\Large $\bullet$}). Idem in $S_3$: $N=1000$ ($\blacklozenge$) and $N=5000$ ($\blacksquare$). \\
The quenching rate is equal to $10^{11}$K/s.
}
\label{Fig3}
\end{figure}

\end{document}